\documentstyle[aps,prl,epsf,twocolumn,epsfig,floats]{revtex}

\begin{document}
\draft



\wideabs{

\title{Electrons on the double helix: optical experiments on native DNA}
\author{E. Helgren, A. Omerzu and G. Gr\"{u}ner}
\address{Dept. of Physics and Astronomy, University of California Los Angeles, Los Angeles, CA 90095}
\author{D. Mihailovic, R. Podgornik}
\address{Josef Stefan Institute, Jamova 39, SI-1000 Ljubljana, Slovenia, and
Department of Physics, University of Ljubljana, Jadranska 19, SI-1000 Ljubljana, Slovenia}
\author{H. Grimm}
\address{Institut f\"{u}r Festk\"{o}rperforschung, Forschungszentrum J\"{u}lich GmbH, D-52425 J\"{u}lich, Germany}

\date{\today}
\maketitle
\begin{abstract}

Optical experiments on calf thymus DNA films subjected to different buffer environments are
reported. The optical conductivity is that of a disordered or lightly doped semiconductor with a 
well-defined band-gap for charge excitations and low frequency transport determined by a small number 
of strongly localized electron states.

\end{abstract}
} 

The question whether electrons are delocalized along the DNA duplex has
attracted substantial recent interest and controversy. Both short and
long-range migration was found by studies that were chemical in nature\cite
{Lewis}. DC conductivity measurements on single DNA strands led to equally
controversial results - with metallic\cite{Fink}, semiconducting\cite{Porath}
and insulating\cite{Braun,de Pablo} behavior all observed. Recent
experiments \cite{Kasumov} are also suggestive of a proximity-induced
superconductivity at low temperatures. DNA networks in contrast were found
to be highly resistive\cite{Okahata}. Our earlier experiments\cite{Tran}
conducted on $\lambda $-DNA in the microwave spectral range gave evidence
for a large resistance associated with the duplex and for a thermally driven
transport process.

These conflicting results may, to a large extent be due to different DNA
varieties. A native DNA duplex with random (or nearly random) base pair
sequences is expected to have electron states that are different than an
oligomer with identical base pairs, such as a poly(C)-poly(G) track. In
addition, the DNA configuration is sensitive to the buffer environment,
which surrounds the duplex. In a dry form the duplex has non-parallel base
pairs, and at higher water concentrations the duplex undergoes a structural
change to a more ordered system\cite{Pohl}. This then may lead to widely
different transport properties and also charge excitations of a different
nature.

In order to examine the nature of electron states in native DNA, we have
conducted optical measurements spanning a wide spectral range, from
microwave frequencies to the UV part of the electromagnetic spectrum, on
oriented films, fabricated from DNA extracted from calf thymus. Our findings
concerning the vibrational modes will be discussed elsewhere. Here we focus
on the electronic excitations at high and at low energies. Our data gives
evidence for well-defined charge excitations at high energies involving the
p-orbitals of the base pairs, and at low energies charge excitations
displaying all the characteristics of conduction due to a small number of
localized electron states. Thus our experiments suggest that DNA is a wide
bandgap semiconductor, with intrinsic disorder, counterion fluctuations, and
possibly other sources leading to a small number of localized electronic
states on the base pair sequence.

Wet-spun, free-standing oriented samples were prepared from calf thymus
Li-DNA (Pharmacia) with a molecular weight of $10^{7}$ (corresponding to
contour length of 5 mm or some 100 persistence lengths) by a method
described by Rupprecht et al\cite{Rupprecht}. This spinning allows
controlled production of sufficient amounts of highly oriented thin films by
spooling DNA fibers that are continuously stretched during precipitation
into an aqueous alcohol solution. Films of thickness of 0.2 mm and lateral
dimensions between 5 and 10 mm were used.

In all applicable cases placing it in a chamber with appropriate relative
humidity controlled the hydration of the DNA sample. Several configurations
were employed to evaluate the optical conductivity $\sigma _{1}\left( \omega
\right) $. In all cases we treat the collection of DNA strands as a
collection of thin wires, of diameter 2 nm, and we define the conductivity $%
\sigma $ as $j/E$ where $j$ is the electric current density induced along
the helix axis. For randomly coiled DNA strands the loss due to motion of
electric charges $W$ is, to a good approximation, given by $W=\frac{1}{3}%
V\sigma \left( E_{0}\right) ^{2}$, where $V$ is the volume of the sample and 
$E_{0}$ is the time averaged applied ac field at the position of the sample.
At 12 GHz the conductivity was evaluated from the measured loss of highly
sensitive resonant cavities that were loaded with the material. The
technique and the analysis, which leads to evaluation of the conductivity
from the measured losses, is well established\cite{Gruner}. In the
millimeter spectral range, 100 GHz to 1000 GHz, backward wave oscillators
were employed as coherent sources in a transmission configuration\cite
{Schwartz}. The oriented calf-thymus DNA film was placed on a 1 mm thick
sapphire substrate and then held in place by a sheet of 6 $\mu $m thick
mylar to form a three layer system. Transmission as a function of frequency
was recorded in the specified frequency range. In our analysis we utilized
the fact that for plane waves incident normally on a slab of material,
resonances occur whenever the slab is an integer number of half wavelengths.
Thus, using the sapphire as a substrate, resonances occurred approximately
every 50 GHz. Having analyzed the transmission through the sapphire alone
prior to mounting the sample, the optical properties of the substrate were
well characterized. The index of refraction of mylar was taken to be
approximately 1.5, and its extinction coefficient was neglected. Thus using
a three-layer transmission model, each resonance was analyzed for the
optical properties of the DNA film, allowing for a 1.5 $cm^{-1}$ resolution
of the spectrum. Thin tungsten wire grids with a spacing much smaller than
the wavelength of our radiation acted as adjustable polarizers to probe the
sample anisotropy. Transmission measurements in the UV spectral range were
conducted using a Beckman Coulter DU 640 Spectrophotometer.

\begin{figure}[tbh]
\centerline{\epsfig{figure=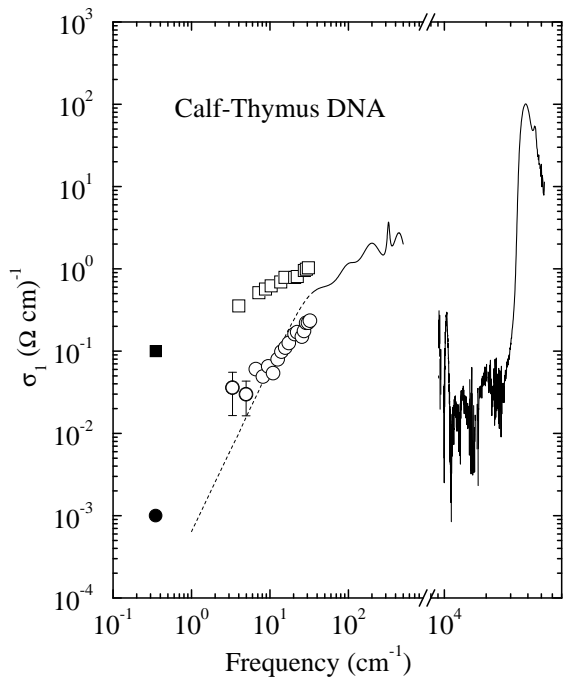,width=8cm}}
\vspace{.2cm} 
\caption{Frequency dependence of the optical conductivity of calf thymus
DNA. Only the conductivity at spectral ranges where electronic excitations
occur are displayed. Vibrational modes (arising between 500 and 10,000 $%
cm^{-1}$) are omitted for clarity.}
\end{figure}

The optical conductivity, measured over a broad spectral range is displayed
in Fig. 1. We have omitted the spectral range between 500 and 10,000 $%
cm^{-1} $, where various intramolecular vibrational excitations occur, which
will be the subject of a separate publication. In the figure we have also
displayed the conductivity extracted from the transmission data of Wittlin
et al.\cite{Wittlin} conducted on dry specimens. Several comments on the
experiments we have conducted are in order. First, during experiments with
an electric field polarized parallel and perpendicular to the duplex axis we
did not observe a substantial anisotropy of the conductivity, and the
absorption peaks associated with vibrational modes were also found to be
isotropic\cite{Wittlin}. While this is surprising, it can be explained by
assuming that the DNA duplex while oriented macroscopically does not assume
a straight configuration but displays a substantial local directional
variation. This has been observed in certain DNA films\cite{Livolant}. As
this issue is unresolved, we have not included a factor of $\frac{1}{3}$ in
the loss equation; this however does not affect our overall conclusion.
Second, it is evident from the figure that data at low frequencies, measured
on dry DNA species is reproducible, that is, the conductivity we have
evaluated is, within experimental error identical to those found by others 
\cite{Wittlin}. Third, the low frequency optical conductivity depends on the
water environment with the conductivity in a wet environment significantly
larger than that in a dry environment. We have found similar results earlier
for $\lambda $-DNA\cite{Tran}.

The mode with the onset at 30,000 $cm^{-1}$ is due to intra-base electronic
excitations associated with the $\pi \rightarrow \pi ^{\ast }$ molecular
orbital transitions. This we have confirmed by conducting optical absorption
measurements on the individual A, T, C and G bases; the absorption as shown
in Fig. 1 is virtually identical to the sum of the optical transitions
associated with the four base pair species. The spectral weight

\begin{equation}
\int \sigma \left( \omega \right) d\omega =\frac{\pi Ne^{2}}{2m}
\label{Eq. 1}
\end{equation}
of the mode, where $N$ is the concentration of charge carriers, and $m$ is
the (electronic) mass, has been evaluated by using the measured extinction
coefficient. By a numerical integration of the spectra for samples of known
dimensions and concentration of DNA, we obtain a value for the concentration
of charge carriers of order $N\sim 10^{21}cm^{-3}$, which compares favorably
with the concentration evaluated by assuming that there is one electron per
base associated with this transition\cite{Dyre}, $N=2\times 10^{21}cm^{-3}$.

On the basis of this analysis we conclude that the absorption feature in the
UV spectral range represents nearly all the spectral weight associated with
the electronic excitations of the base pairs of the DNA duplex. We note that
this excitation cannot be associated with the bandgap in the usual sense, as
this mode does not represent a transition between the highest occupied and
lowest unoccupied states associated with the entire electronic structure of
the DNA duplex. The reason for this is the following: for a duplex, the
optical transition we observe corresponds to a transition between energy
levels of the various single bases, i.e. intra-base excitations, while the
transition matrix element involving energy levels of different bases (such
as A to T or C to G optical transitions) is vanishingly small. The bandgap,
on the other hand, corresponds to the energy difference between the top of
the HOMO band and the bottom of the LUMO band, with these bands in general
corresponding to different bases\cite{de Pablo}.

\begin{figure}[tbh]
\centerline{\epsfig{figure=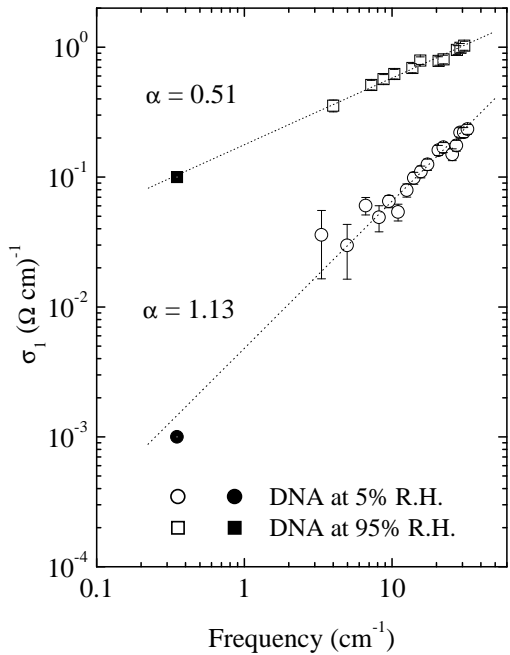,width=8cm}}
\vspace{.2cm} 
\caption{Frequency dependence of the conductivity of calf thymus DNA in a
5\% and a 95\% relative humidity environment. The dashed lines are Eq. (2)
with the parameters given in the figure.}
\end{figure}

The optical conductivity below about 500 $cm^{-1}$, represented by the full
line in Fig. 1 is due to low frequency vibrations involving the double helix
structure. These vibrations have been described earlier\cite{Wittlin} and
are not discussed here. In Fig. 2 the frequency dependent conductivity, as
measured below these vibrational modes in the micro and millimeter wave
spectral range is displayed for both dry and fully hydrated DNA. The dashed
lines represent a conductivity given by the expression

\begin{equation}
\sigma _{1}\left( \omega \right) =A\left( T\right) \omega ^{\alpha }
\label{Eq. 2}
\end{equation}

with the exponent $\alpha $ given in the figure. Such a power law dependence
has been observed in a variety of disordered solids, ranging from ionic
glasses\cite{Livolant} to materials where the electron states are localized 
\cite{Carini}. We believe that contributions to the conductivity due to the
counter ions and water molecules, which surround the DNA duplex, can be
neglected, and that the conductivity is due to localized electrons or holes,
for several reasons. First, both the counter ions and the water in the
hydration layers are strongly bound to the (negatively charged) DNA duplex
and consequently excitations due to these should occur only at higher
frequencies in the infrared spectral range and above. Second, the number of
counter ions does not vary during the hydration, and thus they cannot be
responsible for the strong increase of the conductivity with the increasing
water content. The increased conductivity measured for the hydrated DNA may,
in principle, originate from the response of the water molecules surrounding
the duplex in the hydration layer. We believe this unlikely as we have
measured the dependence of the optical conductivity as a function of
humidity, and have found a strongly nonlinear dependence of said
conductivity on the number of water molecules surrounding the duplex.
Experiments on frozen samples\cite{Warman} of the same composition, where
water molecules and counter-ions are immobilized, show similar behavior
indicating that the contribution of the counter-ions to the optical
conductivity is negligible. A more likely origin of the difference between
the conductivity of the dry and water saturated DNA is as follows. In a
water rich or high humidity environment, the DNA duplex takes on a more
ordered structure (bases stacked parallel to each other) referred to as its
B-form as compared to when the DNA duplex is in a dry or low humidity
environment, where the structure is less ordered (bases are stacked with
different angles with respect to the main axes of the molecules) and
referred to as its A-form. We believe that the increased conductivity is due
to the more ordered arrangement of the DNA system in its B-form as compared
to the disordered A-form allowing for increased electronic charge transport
in spite of the fact that distances between neighboring bases are shorter in
the A-form. In Fig. 3 we display the temperature dependence of the
conductivity measured at 12, 150 and 500 GHz as a function of temperature.
Several features are of importance. First we find a strongly temperature
dependent conductivity, and the temperature dependence is typical of a
conductivity $\sigma _{1}$ determined by temperature driven transport
processes. Second, the temperature dependence itself depends only weakly on
the frequency over a substantial frequency range.

The temperature and frequency dependence of the conductivity as observed at
low frequencies has all the hallmarks of a transport process that is
determined by transitions between localized electronic states. The frequency
dependence of the conductivity under such circumstances is well described by
Eq. (2) with the exponent $a$ depending on the conduction process. For
instance, in the case of variable range hopping (VRH) the strength of the
electron-electron interactions can change the exponent $a$. For VRH, in the
case of non-interacting electrons one finds an approximately linear
dependence of the conductivity on frequency whereas in the case of
interacting electrons, one finds a quadratic dependence of the conductivity
on frequency (plus logarithmic corrections)\cite{Efros & Shklovskii}. We
also find a conductivity that is well described by Eq. (2) with exponents
different for dry and water saturated DNA. The reason for this difference is
not clear; nevertheless the frequency dependence observed is clearly similar
to that of a random assembly of charged entities. In the parameter region $%
k_{B}T>\hbar \omega $, the prefactor $A\left( T\right) $ in Eq. (2), should
display a non-exponential temperature dependence. The function $A\left(
T\right) $ depends on the overall energy scale $E_{0}$\ associated with the
localized states. Variable range hopping is the dominant dc conduction
mechanism if $E_{0}>k_{B}T$, for which

\begin{equation}
\sigma _{DC}=B\exp \left[ -\left( \frac{T_{0}}{T}\right) ^{\beta }\right]
\label{Eq. 3}
\end{equation}
in one dimension, with $\beta =%
{\frac12}%
$, no matter the strength of electron-electron interactions, and the
characteristic temperature $T_{0}$ depending on the localization length and
also on the density of states of the carriers. The temperature dependence is
non-exponential, and in addition becomes progressively weaker with
increasing frequency. The measured temperature dependence displayed in Fig.
3 is in qualitative accordance with such behavior. This then strongly
suggests states localized by disorder. Both static disorder associated with
the random base pair sequences, and also fluctuations involving the DNA
duplex may be responsible\cite{Bruinsma} for determining the overall
temperature and frequency dependence.

\begin{figure}[tbh]
\centerline{\epsfig{figure=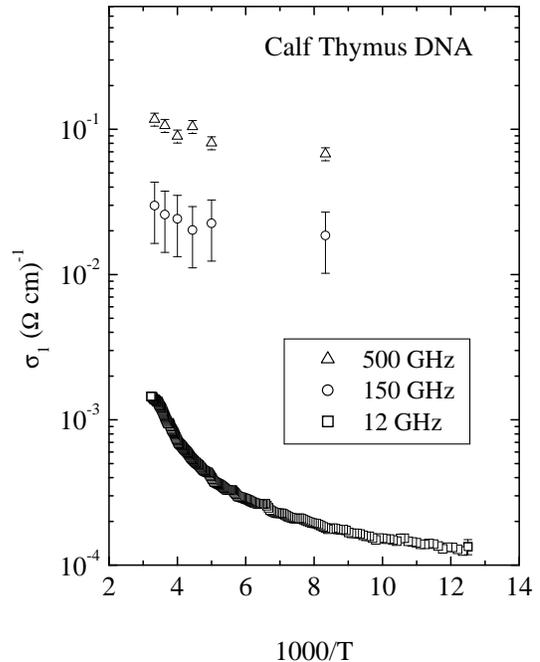,width=8cm}}
\vspace{.2cm} 
\caption{Temperature dependence of the conductivity of dry calf thymus DNA
measured at 12, 100 and 500 GHz.}
\end{figure}

Our experimental results, obtained over a broad spectral range strongly
suggest that native DNA is a wide bandgap semiconductor, with disorder and
counterion fluctuations (among other possible sources) leading to a small
number of localized carriers on the base pairs - a situation not unlike what
occurs in lightly doped crystalline or amorphous semiconductors or doped
polymers\cite{Menon}. Under such circumstances two important contributions
to the optical conductivity emerge: a well defined transition associated
with the electron states of the ''pure'' system (this transition occurring
at finite energy) together with a low frequency contribution to the
conductivity with prominent frequency and temperature dependencies. We
observe both features in native DNA with some modifications from what is
found for a typical lightly doped (crystalline or amorphous) semiconductor
or polymer. In the latter case one observes a well-defined optical
transition between the valence and conduction band, as discussed earlier,
for DNA, the transition is between orbitals associated with individual
bases. Thus the fact that band structure calculations\cite{de Pablo} give a
bandgap smaller than that corresponding to the energy for the onset of
absorption in Fig. 1 is not surprising.

The number of carriers involved in the low frequency transport process is
difficult to estimate as the parameters that enter into the equations for
the temperature and frequency dependent conductivity are determined by both
the number of carriers, and by the localization length. A comparison with
other disordered one-dimensional conductors however allows one to draw some
conclusions and to make some order of magnitude estimates. The temperature
dependence of the conductivity that we find for DNA is similar to a strongly
disordered organic conductor $Qn(TCNQ)_{2}$ with disorder induced by
irradiation\cite{Holczer}, for which the localization length is $\xi \sim 10$%
\AA , and the number of carriers is $N\sim 10^{22}cm^{-3}$. The magnitude of
the conductivity we find here is approximately three orders of magnitude
smaller than that observed in $Qn(TCNQ)_{2}$. For a conduction process
determined by temperature driven transitions between localized electron
states, the localization length, $\xi $, mainly determines the overall
temperature dependence. The magnitude of $\sigma _{1}$ reflects both $\xi $
and the number of carriers, $N$. Thus the comparison between calf-thymus DNA
and $Qn(TCNQ)_{2}$ suggests that in native DNA electron states are
characterized by a short localization length not exceeding one lattice
constant (the distance between base pairs), while the number of carriers
participating in the low frequency transport process is of the order of $%
N\sim 10^{19}cm^{-3}$. The higher conductivity of DNA in a water rich
environment may reflect a different carrier number, but most likely it is
due to the more regular B-form of DNA which occurs for high water content,
thus also leading to weaker localization effects, and thus to a larger
localization length.

The source of the charge entities (electrons or holes) which are
associatedwith the low frequency conductivity is not obvious. The random
base pair sequences, together with disorder associated with the finite
persistence length, and counter ion fluctuations all may lead to a small
number of localized charges on the base pair stack. The fact that the
states, which contribute to the low frequency optical response, are
localized has different ingredients. The random base pair sequences which
occur in native DNA lead to a random potential along the duplex, and thus to
charge localization. The potential energy fluctuations associated with the
base pair sequences are of the order of $0.5eV$, about ten times larger than
the overlap integral between the electron states on neighboring base pairs
along the double helix\cite{Bruinsma,Jortner}. Under such circumstances the
localization length is of the order of one lattice constant, comparable to
the value we inferred before. One should also note that equally important is
the dynamic disorder associated with base pair fluctuations, of which the
influence on charge transport along the DNA duplex has been conjectured
earlier\cite{Bruinsma}.

We wish to thank Phu Tran for assisting with the cavity measurements. This
research was supported by the National Science Foundation grant DMR-0077251.

\end{document}